\documentclass[12pt,referee,sn-aps,Numbered]{sn-jnl-custom}

\usepackage{etoolbox}
\makeatletter
\patchcmd{\@afterheading}{\@afterindenttrue}{\@afterindenttrue\@nobreaktrue}{}{}
\makeatother

\usepackage{graphicx}% Include figure files
\usepackage{dcolumn}% Align table columns on decimal point
\usepackage{amssymb}
\usepackage{bm}% bold math
\PassOptionsToPackage{hyphens}{url}\usepackage{hyperref}% add hypertext capabilities
\usepackage{tabularx}
\usepackage{float}
\usepackage[dvipsnames]{xcolor}
\usepackage[T1]{fontenc}	% should generally be included for better accented-word behavior
\usepackage[latin9,utf8]{inputenc}	% should generally be included for better accent behavior
\usepackage{geometry}

\usepackage{comment}
\hypersetup{colorlinks=true,urlcolor=blue,citecolor=blue,linkcolor=blue}   
\usepackage[T1]{fontenc}	% should generally be included for better accented-word behavior
\usepackage[latin9,utf8]{inputenc}	% should generally be included for better accent behavior
\usepackage{graphicx}
\usepackage[above,below]{placeins}	% allows use of \FloatBarrier command to force section barriers
\usepackage{multirow}
\usepackage[dvipsnames]{xcolor}
\usepackage{ifthen, scalerel}
\usepackage{makecell}
\usepackage{subcaption}
\expandafter\let\csname equation*\endcsname\relax \expandafter\let\csname endequation*\endcsname\relax

\usepackage{amsmath} 
\usepackage{stringstrings}

\newcommand{\ket}[1]{|#1\rangle}

\newcommand{\fracroot}[2]{\ifthenelse{#1=1}{\frac{1}{\sqrt{#2}}}{\sqrt{\frac{#1}{#2}}}}

\usepackage{makecell}

\usepackage{babel,csquotes,xpatch}
\usepackage[utf8]{inputenc}

\makeatletter
\renewcommand\@makefnmark{\hbox{\@textsuperscript{\normalfont\arabic{footnote}}}}
\makeatother

\newcommand\sbullet[1][.5]{\mathbin{\vcenter{\hbox{\scalebox{#1}{$\bullet$}}}}}

\makeatletter
\renewcommand\@makefntext[1]{%
  \noindent
  \hbox{\@textsuperscript{\normalfont\arabic{footnote}}}\,#1%
}
\makeatother

\makeatletter
\let\@fnsymbol\@arabic
\makeatother
    
\begin{document}

\title[]{Assessing student learning in quantum computing: Lessons from developing a test item on phase kickback}

\author*[1,2]{\fnm{Josephine C.} \sur{Meyer}}\email{jcmeyer26@csusm.edu}

\author[3]{\fnm{Steven J.} \sur{Pollock}}\email{steven.pollock@colorado.edu}

\author[3]{\fnm{Bethany R.} \sur{Wilcox}}\email{bethany.wilcox@colorado.edu}

\author[4]{\fnm{Gina} \sur{Passante}}\email{gpassante@fullerton.edu}

\affil*[1]{\orgdiv{Dept.\ of Physics}, \orgname{California State University San Marcos}, \orgaddress{\street{333 S Twin Oaks Valley Rd.}, \city{San Marcos}, \postcode{92096}, \state{CA}, \country{US}}}

\affil[2]{\orgdiv{Dept.\ of Physics and Astronomy}, \orgname{George Mason University}, \orgaddress{\street{4400 University Drive, MSN: 3F3}, \city{Fairfax}, \postcode{22030}, \state{VA}, \country{US}}}

\affil[3]{\orgdiv{Dept.\ of Physics}, \orgname{University of Colorado Boulder}, \orgaddress{\street{390 UCB}, \city{Boulder}, \postcode{80309}, \state{CO}, \country{US}}}

\affil[4]{\orgdiv{Dept.\ of Physics}, \orgname{California State University Fullerton}, \orgaddress{\street{800 N State College Blvd.}, \city{Fullerton}, \postcode{92831}, \state{CA}, \country{US}}}

\abstract{
A major challenge for quantum workforce development is the need to both understand and reliably assess student learning of quantum information science (QIS) fundamentals. Yet student thinking is notoriously difficult to probe, even for seasoned education researchers.
This article presents the story of Item 15 on the Quantum Computing Conceptual Survey (QCCS). This assessment item underwent more revision and discussion within the team than the remaining 19 assessment questions combined. This paper provides a behind-the-scenes look at the development of this assessment question: a story that both reveals interesting findings about student reasoning in quantum computing and illustrates why quantum education researchers insist on triangulating diverse quantitative and qualitative data sources when developing and refining assessment items, with implications for any researcher looking to understand and measure student conceptual reasoning in quantum computing, as well as for QIS curriculum and workforce development more broadly.}


\maketitle

\section{Introduction}
In recent years, researchers from various discipline-based education research (DBER) communities have come together to study student thinking in quantum technologies \cite{Meyer:2026AJP}, with the goal of developing better curricular materials, pedagogies, and assessments. This work is often disseminated through venues that reach a broad audience of scientists, engineers, education researchers, and educators. Because these audiences have varying exposure to education research, communication barriers can arise that limit accessibility and the broader impact of quantum education results. This article aims to help bridge a portion of that gap -- demonstrating, by way of example, how we as education researchers come to understand aspects of student thinking, and motivating some of the methodological and practical considerations behind our work in ways we hope will be accessible and useful to a broad audience.

One of the primary goals of quantum information science (QIS) education research is to study how students think about QIS principles. A major challenge, however, is that unlike much experimental quantum science research, education researchers can seldom conduct controlled experiments. Even innovative techniques like eye-tracking \cite{Hahn:2022} provide at best indirect evidence of student thinking.
Instead, we gather data about student thinking in several common ways: for instance, by looking at students' work on problems, observing them in lectures and tutorials \cite{Schoenfeld:2018}, and performing think-aloud interviews \cite{Leighton:2017}. Each data collection method has strengths and limitations. For instance, clinical interview dynamics affect student interviews \cite{Sherin:2007,Russ:2012}, while submitted problem sets sometimes present a ``cleaned-up'' narrative erasing the thinking behind the solution. Qualitative analyses are also difficult to scale, motivating development of quantitative tools such as research-based conceptual assessments \cite{Madsen:2017} to complement (but not replace) qualitative research. =

Even with these methods, student thinking remains challenging to understand, particularly in a new and interdisciplinary field like QIS. 
The best way to illustrate this is by example. In response to calls from quantum workforce development experts and advocates \cite{Aiello:2021}, our team sought to develop a research-based assessment for quantum computing \cite{Meyer:2025QCCS}, necessitating closed-response questions that would accurately measure students' understanding. The development of these questions is iterative, as edits are made between versions to clarify the question or hone in on different reasoning options. 
In this article, we discuss and analyze the development of one specific assessment item, whose story
both reveals insight about student thinking (Sec.~\ref{sec:discussion}, see also Ref.~\cite{Plueger:2026})  and illustrates the necessity of triangulating multiple data sources to ensure that a multiple-choice test item actually measures the conceptual reasoning pathways we intend to measure.

\subsection{Quantum Computing Conceptual Survey}

As part of a larger project on student thinking about quantum computing, we have developed a closed-form, multiple-choice conceptual assessment called the Quantum Computing Conceptual Survey (QCCS) \cite{Meyer:2025QCCS,Meyer:2025Thesis}. We intend for QCCS to provide a reliable means for assessing student learning across diverse course contexts. This instrument was informed by instructor perspectives \cite{Meyer:2022PhysRev} and early studies of student reasoning in quantum computing contexts (e.g.\ \cite{Passante:2015,Wan:2019,Meyer:2021PERC,Meyer:2022PERC,Kushimo:2023}). Content coverage was defined and constrained via a survey of faculty teaching introductory quantum information science courses, selecting only topics assessed by at least 80\% of the $N=68$ surveyed faculty \cite{Meyer:2024EPJ}. The instrument has undergone iterative cycles of validation and refinement based on both statistical analysis as well as student and instructor interviews \cite{Meyer:2025QCCS}. Note that this paper is \textit{not} intended to be a detailed argument for the validity and reliability of the QCCS; please instead refer to Ref.~\cite{Meyer:2025QCCS}.

\subsection{Understanding phase kickback}

The QCCS item we focus on in this paper centers ideas critical to phase kickback, though we anticipate the insights discussed should be broadly applicable across QIS education research. Phase kickback is a conceptually challenging element in many quantum computing algorithms with no straightforward analog in classical computing \cite{Kaye:2007}. 
While the reader need not fully understand phase kickback to appreciate the gist of this article, we provide a brief overview for those interested. 

\subsubsection{CNOT gate}

To understand phase kickback, we must first define the CNOT gate. CNOT is a 2-qubit entangling gate that, together with single-qubit rotations, forms a universal gate set for quantum computation. It is denoted:

\begin{center}
    \includegraphics[height = 4em]{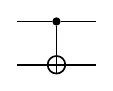}
\end{center}
with the black dot denoting the ``control'' qubit and the white cross denoting the ``target.'' For this paper, we will always treat the first/top qubit as the control qubit, reflecting how students encounter the CNOT gate in the assessment item discussed here. 

When the control qubit is in the $\ket0$ state, the CNOT gate has no effect and acts as the identity operator. When the control qubit is in the $\ket1$ state, the CNOT gate applies a rotation to the second qubit mapping $\ket0 \leftrightarrow \ket1$.
In general, with respect to the standard basis states, the CNOT gate behaves like a classical XOR gate, where the bottom bit is the result of the XOR and the top bit is copied:
\begin{gather}
\ket{00}\rightarrow\ket{00} \notag\\
\ket{01} \rightarrow\ket{01} \notag\\
\ket{10} \rightarrow\ket{11} \notag\\
\ket{11} \rightarrow\ket{10} \notag
\end{gather}
However, the classical XOR analogy ends there. When a superposition state is input, CNOT produces entanglement -- a phenomenon for which there is no classical analog. For instance, the states $\ket{\pm}\otimes\ket0$ are mapped to/from the entangled Bell states $\ket{\Phi^{\pm}} \equiv \fracroot{1}{2}(\ket{00} \pm \ket{11})$. In entangled states, \textit{neither qubit's state can be written as a well-defined ket}
.\footnote{

The state of individual qubits in an entangled state \textit{can} be written as a mixed state using density matrices, but this formalism is seldom taught in undergraduate quantum computing courses \cite{Meyer:2024EPJ} and is therefore avoided throughout the QCCS.
}

\subsubsection{Phase kickback}

Even when entanglement is not generated, the CNOT's ``control'' and ``target'' labels can become misleading. For example, the CNOT gate maps the states $\ket\pm\otimes\ket-\leftrightarrow\ket\mp\otimes\ket-$. In this instance, the state of the ``target'' qubit is unchanged, yet the state of the ``control'' qubit \textit{does} change. The term ``phase kickback'' is given to this mapping and comes from the phase of the target qubit being ``kicked back'' to the phase of the control qubit. More generically, in this paper, we say that phase kickback occurs whenever CNOT changes the state of the control qubit.

Phase kickback is central to many textbook quantum algorithms (e.g.\ Deutsch-Jozsa \cite{Deutsch:1992}, Berstein-Vazirani \cite{Bernstein:1993a}, and Shor's factoring algorithm \cite{Shor:1994}), and was identified by Kushimo and Thacker \cite{Kushimo:2023} as a source of difficulty for students. For these reasons, it was desirable to include a question on phase kickback on the assessment, as long as doing so was compatible with the broader goals of the instrument. 

\subsection{Our assessment objectives}

When designing an assessment, developers first
articulate measurable \textit{assessment objectives} that can be mapped to multiple items \cite{Vignal:2022}, ideally representing overarching, context-independent concepts or skills that students can master. Test items are mapped to specific objective(s), aiming to cover all objectives within a reasonable number of items.

We developed 21 assessment objectives, grouped according to 5 overarching concepts, based on the initial topics survey \cite{Meyer:2024EPJ} and refined based on input from $N=6$ instructor interviews. Items were crafted from these 21 assessment objectives in an iterative process.\footnote{
In practice, given the constraints of semester piloting timelines, we developed the initial test items and assessment objectives simultaneously through an iterative process. We only present the final objectives here. 
} 
Item 15 implemented 4 objectives (Table~\ref{tab:AOs}).

\begin{table}[!h]
    \centering
    \begin{tabular}{>{\raggedright\arraybackslash}p{150pt} p{300pt}}
        \hline \hline
        \makecell{\thead{Overarching concept} \\ \ }
         &  \makecell{\thead{Assessment objectives relevant to Item 15} \\\textit{Students should be able to ...}} \\
         \hline
         1. Mathematical foundations & --\\
         2. Entanglement & \textit{2B. Given a multi-qubit state, find the state of a subset of qubits (or state that it is not well-defined as a ket).} \\
        3. Superposition and measurement & \textit{3B. Write the state of a 1- or 2-particle superposition state before and after measurement (including partial measurement), or state that it is ill-defined.}\\
        4. Quantum gates & \textit{4A. Evaluate the eﬀects of standard gates (I, X, Z, H, CNOT) on 1- and 2-qubit quantum states.} \\
        5. Quantum circuit diagrams & \textit{5A. Compute the final state of a 1- or 2-qubit quantum circuit given a starting state and sequence of gates.} \\
         \hline \hline
         
    \end{tabular}
    \caption{QCCS assessment objectives upon which Item 15 was based. Note that while none of these objectives explicitly target phase kickback (a topic too specific to incorporate directly into assessment objectives), this combination of 4 objectives was highly amenable to a question on phase kickback.}
    \label{tab:AOs}
\end{table}

\subsection{Designing high-quality multiple-choice items}

To promote faculty uptake, we wanted QCCS to eventually feature only closed-form, machine-gradable items. Thus, the QCCS features a mixture of dichotomously-scored multiple-choice and multiple-response items. In a few cases, related questions sharing a single stem are scored as a single item.\footnote{

Dichotomous scoring was mandated by our intention to utilize the Rasch model, alongside the more commonly-used classical test theory (CTT), for QCCS validation \cite{Meyer:2025Thesis} -- a decision motivated by the stronger metrological foundation of Rasch measurement theory \cite{Ding:2023} and the highly heterogeneous academic backgrounds of students in introductory quantum computing classes \cite{Meyer:2022PhysRev}. Rasch analysis with non-dichotomous items requires significantly larger datasets. The Rasch model also assumes items are structurally independent, motivating treating multiple questions with the same stem as part of the same item. For this paper, however, we stick to CTT as Rasch analysis revealed qualitatively similar findings for Item 15.

}

Designing high-quality multiple-choice test items is both a science and an art 
(e.g.\ Ref.~\cite{Engelhardt:2009}). To meet the standards of rigor of a research-based assessment, each item ought to (a) align with the targeted assessment objectives, (b) behave acceptably by statistical benchmarks, (c) perform well in instructor and student interviews, and (d) produce a sensible and interpretable pattern of incorrect answers.
Only items that pass all four criteria are deemed suitable for the final assessment.

Generally, strong test items share a few common features \cite{Engelhardt:2009}. Items should be straightforward and concise, avoiding unnecessary ``gotchas,'' roundabout language (e.g.\ double negatives), or culturally-specific references. Items should be insensitive to competing notational and terminology conventions, and resistant to test-taking tricks and haphazard guessing. Ideally, items should incorporate tempting ``distractors'' based on known naive conceptions or reasoning errors, to lure low-performing students away from the correct answer. 

While quantitative metrics can flag problematic test items, it is often much more difficult to figure out \textit{why} a test item is misperforming. Diagnosis requires triangulation of multiple methodologies (quantitative and qualitative), and often much trial-and-error. Even for experienced assessment developers and education researchers, reasons for item failure frequently elude us. This returns us to the primary point of this paper: student thinking is challenging to measure, and seemingly minor changes to problem wording can dramatically impact the reasoning students use.\footnote{This phenomenon is by no means limited to the context of QIS education; however, the formative stage and interdisciplinary nature of QIS likely only exacerbates these challenges compared to more-established fields like undergraduate physics.} Accordingly, lessons learned from troubleshooting underperforming test items are often as valuable as the final assessment scores.

\section{Context}
In this work, we reflect on the development of one specific QCCS question (Item 15).  We will describe the versions of the question, data sources (quantitative and qualitative) used to evaluate the item's performance, and refinements made at each iteration. Quantitative data was collected from large-scale administrations of the survey to students across the United States \cite{Meyer:2025QCCS}. All data collection was via informed consent and approved by the University of Colorado Boulder institutional review board.
A timeline of the item development process is shown in Table~\ref{tab:i15-flowchart}.

\begin{table}[!h]
    \centering
    \begin{tabular}{c | c c c c c}
        \hline \hline
        & \thead{Pilot round} & \thead{QCCS v1.0} & \thead{QCCS v2.0} & \thead{QCCS v2.1} & \thead{QCCS v2.2} \\
        \hline

        & \thead{Semesters} & Fall 2023 & Sp/Sum 2024 & Fall 2024 & Sp-Fall 2025       \\
        \hline
  \hspace{-5mm}\parbox[t]{4mm}{\multirow{6}{*}{\rotatebox[origin=c]{90}{\thead{Data collected}}}} \\
  
        & \textit{Student responses} & 271 & 621 & 346 & 777 \\

        & \textit{Think-aloud interviews} & 6 & 30 & -- & -- \\
        & \textit{Expert interviews} & 6 & -- & -- & -- \\
        & \textit{Open-response version} &  $\checkmark$ & -- & -- & --
         \\    
        & \textit{Free-form feedback box} &  $\checkmark$ & $\checkmark$ & -- & --
         \\ \\ \hline
        & \thead{Change log} & -- & \textit{\makecell{$\sbullet$ Specific input state \\
        $\sbullet$ Graphical time slice \\
        $\sbullet$ ``0/1'' over meter \\
        }} & 
        \textit{\makecell{$\sbullet$ Two-part item \\
        $\sbullet$ Measurement \\ outcome given
        }}
        & \textit{\makecell{$\sbullet$ MC answers \\ revised}
        }\\
        \hline \hline
         
    \end{tabular}
    \caption{Timeline of QCCS development and validation, focusing on key data sources for the refinement of Item 15. Change log is elaborated in subsequent sections. Check mark indicates the data was collected for the given pilot round.}
    \label{tab:i15-flowchart}
\end{table}

\subsection{Statistical metrics}

Statistical validation of PER assessments is a complicated process. For simplicity, we limit our discussion to two figures of merit that proved important for troubleshooting Item 15 specifically:

\begin{itemize}
    \item \textbf{Item difficulty ($p_i$):} The fraction of pilot students who answer item $i$ \textbf{correctly}. Generally, the range $p \in [0.3, 0.9]$ is preferred to avoid saturation \cite{Engelhardt:2009}. Harder items may be acceptable if the guess floor is sufficiently small.
    \item \textbf{Item discrimination:} An item's ability to distinguish between low- and high-achieving students. We use a modified point-biserial $\rho_i^*$ that measures Spearman correlation between item $i$ and overall score on the remaining items. We aim for discrimination $\rho^*\ge0.2$ \cite{Engelhardt:2009}; larger values (i.e. $\rho^*\ge0.3$) are preferred. Very easy/hard items tend to have lower discrimination.
\end{itemize}

For each iteration, we also report response frequencies (correct and by distractor) as a function of binned overall score on the remaining items. Such curves are a useful diagnostic heuristic for troubleshooting malfunctioning items; they do not constitute a statistical claim or figure of merit in themselves.

\section{Results and analysis}
Each subsection below details a single version of this question.
We discuss what we have learned about student thinking in each version and how that led to the creation of the next version.

Where item statistics, e.g.\ difficulty and discrimination, are given for a specific item, values are provided along with bootstrapped standard error \cite{Efron:1994}.\footnote{

Following experimental physics convention, these values are presented as e.g. $p=0.55(2)$ where the value in parentheses represents the $1\sigma$ uncertainty in the last digit, i.e.\ $p=0.55\pm0.02$.

} Further diagnostic information on an item can be extracted from plots of item response vs.\ overall score \cite{Forthmann:2020}. In such plots, student overall score is binned using minimum-deviation quantile assignment under ties \cite{Jackson:2005}.

\begin{figure}
    \centering
    \begin{subfigure}[t]{0.45\textwidth}
        \fbox{\includegraphics[scale = 0.2, clip = true, trim = 50pt 750pt 50pt 800pt]{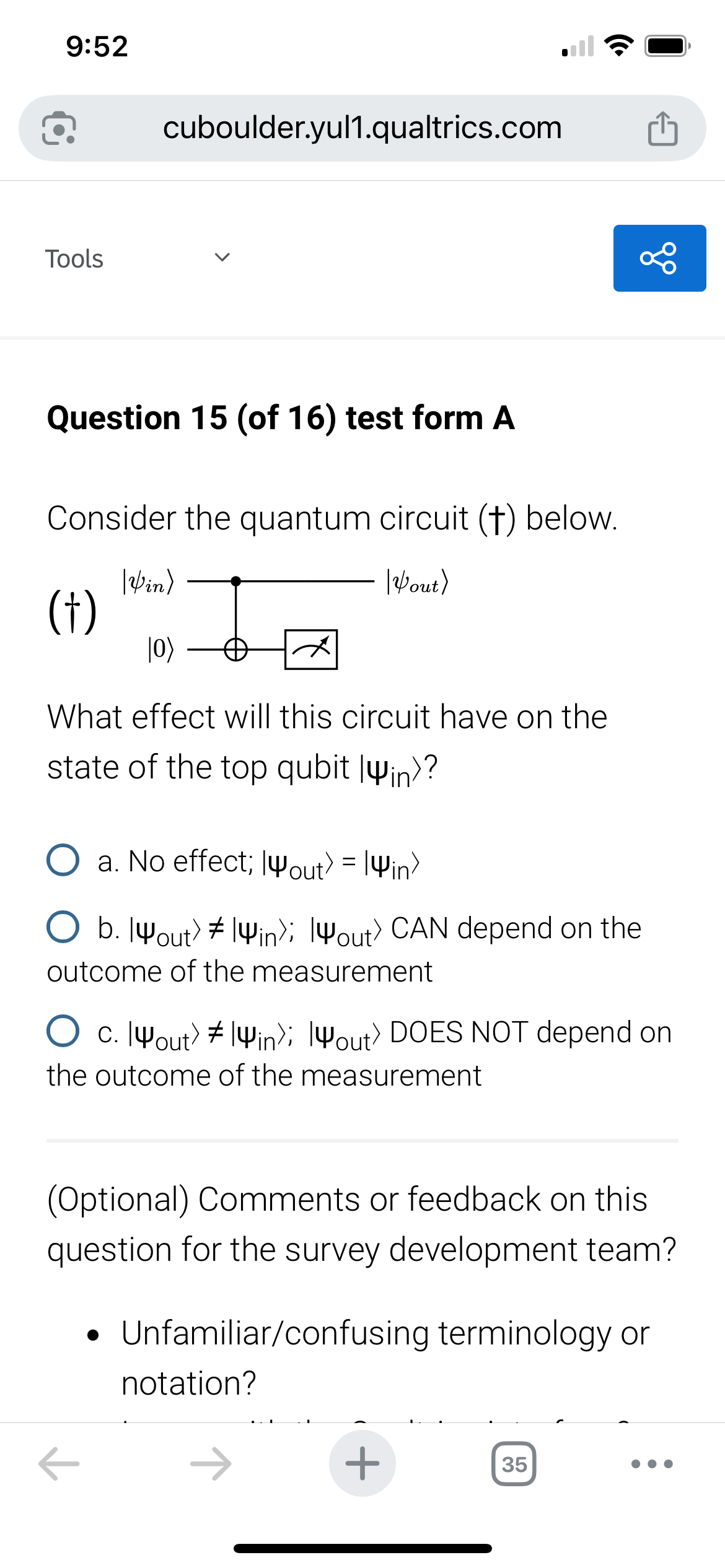}}
        \caption{Item 15 from QCCS version 1.0, variant A. (Correct response: ``b'')}
        \label{fig:v1.0A}
    \end{subfigure}
    \hspace{0.5cm}
    \begin{subfigure}[t]{0.45\textwidth}
        \fbox{\includegraphics[scale = 0.2, trim = 50pt 350pt 50pt 1100pt, clip = true]{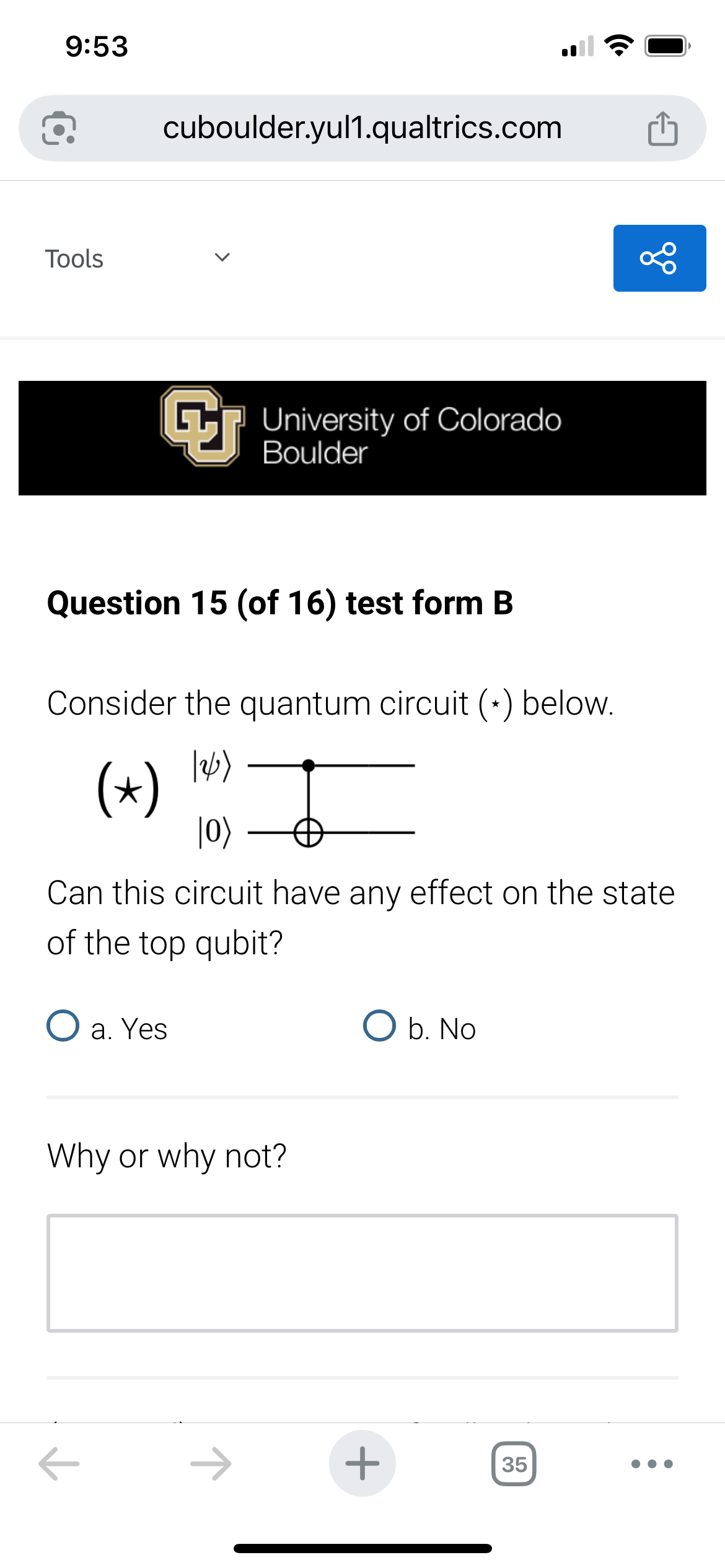}}
        \caption{Item 15 from QCCS version 1.0, variant B. (Correct response: ``a,'' though primary interest here was in open-response box to generate distractors.)}
        \label{fig:v1.0B}
    \end{subfigure}
    \caption{Piloted versions of Item 15 during the initial fall 2023 pilot semester (v1.0).}
    \label{fig:v1.0}
\end{figure}

\subsection{QCCS version 1.0 (fall 2023)}

The first iteration of QCCS focused on generating actionable exploratory feedback from students and instructors (as opposed to large-$N$ statistical rigor). Accordingly, we pilot-tested two different versions of (future) Item 15, one closed-form (Fig.~\ref{fig:v1.0A}) and one open-form (Fig.~\ref{fig:v1.0B}), the latter serving to elicit strong distractor candidates.\footnote{
After completing 8 anchor items, students were randomly assigned to either version A or version B of the assessment, and then had the opportunity to attempt the other version if they wished. Therefore, some students completed both versions, and their performance could be linked.

} We also conducted $N=6$ student think-aloud interviews and $N=6$ faculty interviews with these questions, to identify structural issues to address in subsequent versions. 

These interviews revealed variant B was too unstructured to constitute a viable question. Variant A also showed issues:

\begin{itemize}
    \item In some courses, measurement was implicitly defined to be in the standard ($Z$) basis. In other courses, students were taught that measurement without specifying a basis was a meaningless operation. A notation had to be developed that would cue students in the second camp to interpret the meter as a $Z$ measurement, without introducing the term ``basis'' and confusing students in the first camp.
    \item The word ``effect'' was ambiguous for some students. Is it an effect if the result is to leave the top qubit's state ill-defined as a ket? Is it an effect if the qubit's state has changed, but the change does not alter ($Z$-basis) measurement probabilities?
    \item The word ``CAN'' in answer (b) was likewise a source of confusion for students: did it imply that $\ket{\psi_{out}}$ depended on the outcome of the measurement for every $\ket{\psi_{in}}$, or just for some choices of $\ket{\psi_{in}}$?
    \item Circuit diagrams: Some students struggled to interpret whether the measurement occurred before, after, or simultaneous to the determination of $\ket{\psi_{out}}$ in variant A. Did the $\ket{\psi_{out}}$ label attach to the ``wire'' at its start (after the CNOT gate) or at is finish (where the label was placed)?
\end{itemize}

\subsection{QCCS version 2.0 (spring 2024)}

Based on findings from v1.0, we constructed a largely new version of the assessment for spring 2024 with 20 closed-form items. Item 15 appeared as a single closed-form item, incorporating feedback from faculty interviews, student think-aloud interviews, prior student responses, and responses to optional feedback boxes from v1.0.

\begin{figure}[!htbp]
    \centering
    \begin{subfigure}[t]{0.417\textwidth}
        \fbox{\includegraphics[scale = 0.185, clip = true, trim = 60pt 600pt 85pt 950pt]{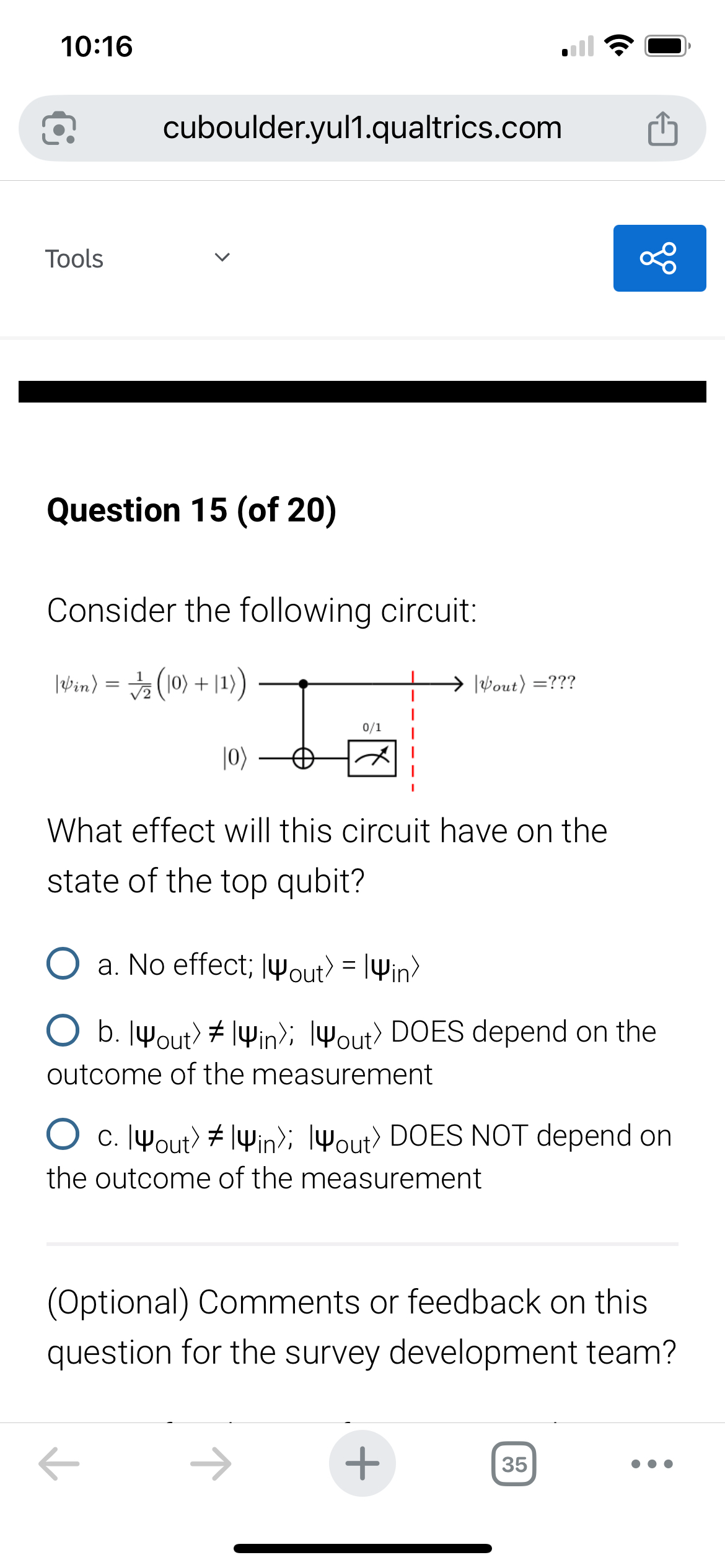}}
        \caption{Item 15 from QCCS version 2.0.}
        \label{fig:v2.0}
    \end{subfigure}
    \begin{subfigure}[t]{0.577\textwidth}
        \fbox{\includegraphics[scale = 0.37, trim = 15pt 750pt 20pt 10pt, clip = true]{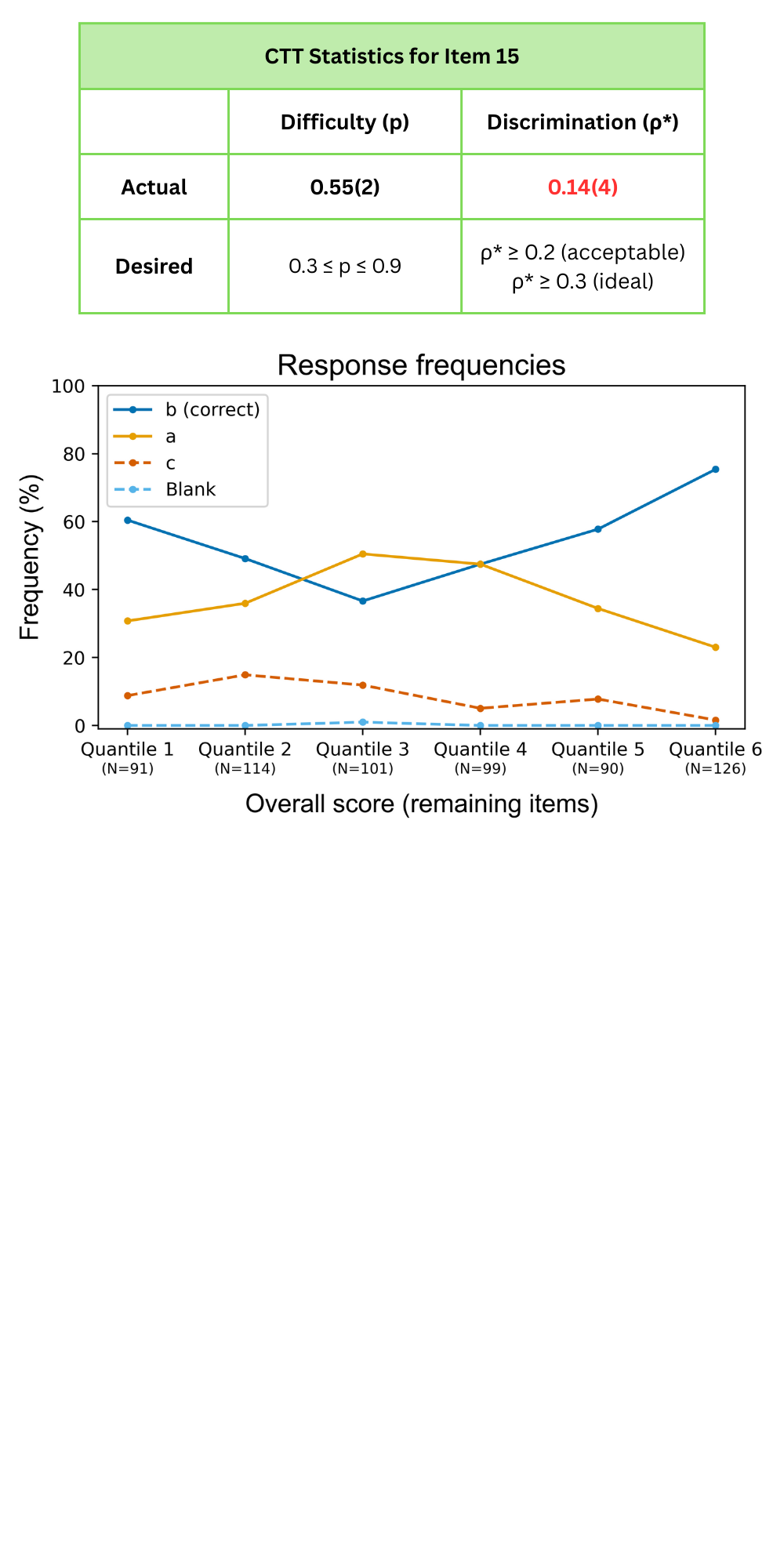}}
        \caption{Item statistics for Item 15, QCCS version 2.0. The plot of distractor frequencies vs.\ binned overall score clearly shows the cause of the low discrimination: the percentage of students choosing the correct answer is essentially flat or decreasing across the lower 2/3 of students.}
        \label{fig:v2.0-stats}
    \end{subfigure}
    \caption{Piloted version of Item 15 during the spring 2024 pilot semester (v2.0), featuring two distinct but parallel subitems and a modified circuit diagram with specified input state.}
    \label{fig:v2.0-with-stats}
\end{figure}

Revisions included:
\begin{itemize}
    \item A specific superposition state, $\ket+$, was given in place of generic $\ket{\psi_{in}}$ to eliminate the word ``can'' and clarify that $\ket{\psi_{in}}$ need not be a basis state.
    \item A dashed vertical line was added to denote a time slice, and an arrow was added on the outgoing wire to clarify that $\ket{\psi_{out}}$ attached to the wire after the time slice.
    \item The ``0/1'' label over the meter (one of several variants beta-tested) proved to be the least ambiguous notation for measurement: students who had only encountered measurement in the $Z$ basis interpreted this notation as indicating (redundantly) a binary bit 0 or 1 was output, while students used to a basis specification treated this as an implicit reminder to use the $Z$ basis as specified in the instructions.
\end{itemize}

The revised version of Item 15 for QCCS v2.0 is shown in Fig.~\ref{fig:v2.0}, alongside performance statistics (Fig.~\ref{fig:v2.0-stats}). For this iteration, we collected $N=621$ viable student responses across 43 courses representing 40 unique US institutions. Despite the revisions above and an excellent item difficulty, this item on v2.0 had unacceptably low discrimination ($\rho^*=0.14(4)$). We then conducted $N=30$ student think-aloud interviews targeting Item 15 and a few other problematic items \cite{Plueger:2026}.

Interviews, along with the response curve shown in Fig.~\ref{fig:v2.0-stats}, revealed that low-performing students often guessed the correct answer ``b'' possibly leveraging multiple-choice guessing strategies. Middle-performing students selected the conceptual distractor ``a'' \textit{more} often than the lowest-performing students, making item performance non-monotonic with overall test score. Interviews also revealed that students could answer correctly without successfully interpreting the state of the top qubit immediately prior to the measurement, prompting us to split the item into two subparts in future iterations.

\subsection{QCCS version 2.1 (fall 2024)}

\begin{figure}[!htbp]
    \centering
    \begin{subfigure}[t]{0.39\textwidth}
        \fbox{\includegraphics[scale = 0.185, clip = true, trim = 50pt 275pt 80pt 700pt]{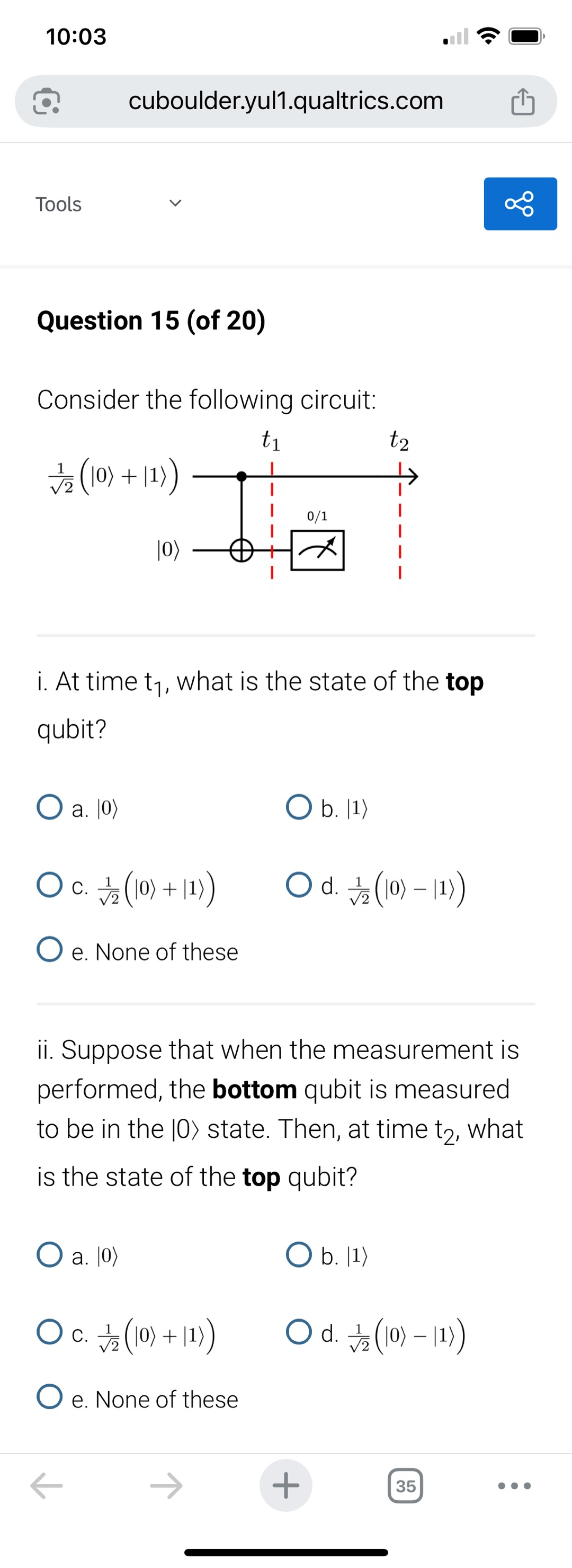}}
        \caption{Item 15 from QCCS version 2.1. Item was scored as correct if student answered ``e'' for part (i) and ``a'' for part (ii); no partial credit awarded if only one subpart answered correctly.}
        \label{fig:v2.1}
    \end{subfigure}
    \hfill
    \begin{subfigure}[t]{0.6\textwidth}
        \fbox{\includegraphics[scale = 0.375, trim = 10pt 100pt 15pt 50pt, clip = true]{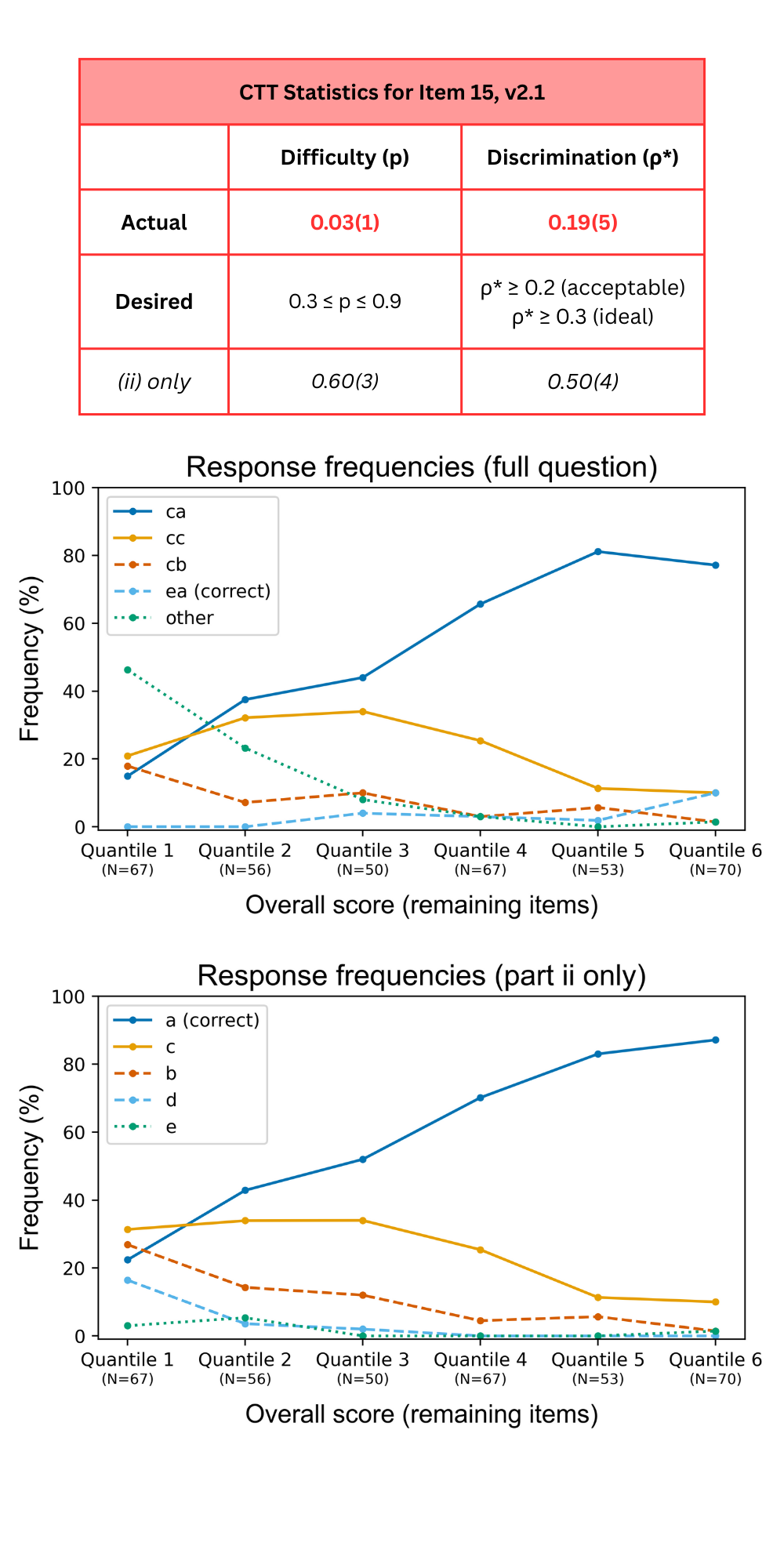}}

        \caption{Item statistics for Item 15, QCCS version 2.1. Observe that part (ii) performs acceptably by itself, but even high-achieving students appeared very reticent to answer ``e'' for part (i).}
        \label{fig:v2.1-stats}
    \end{subfigure}
    \caption{Piloted version of Item 15 during the fall 2024 pilot semester (v2.1).}
    \label{fig:v2.1-with-stats}
\end{figure}

Fig.~\ref{fig:v2.1} shows the revised version of Item 15 for QCCS version 2.1 (fall 2024), developed by consensus of the research team. The new format was intended to address several weaknesses of previous forms of the question:

\begin{itemize}
    \item Subitems now probed student reasoning about the state before \textit{and} after measurement. 
    \item A specified post-measurement state for the lower qubit addressed questions about what was meant by the upper qubit's state ``changing.''
    \item The two-tier question structure lowered the random guessing probability to a negligible 4\%, while the parallel structure of the two subparts' answer choices was chosen to reduce the effectiveness of test-taking strategies.
\end{itemize}

For part (i), we anticipated the correct answer ``e'' would be attractive both to students familiar with density matrices (who could presumably recognize the state as maximally mixed), and those who had not (who could still rule out a-d through process of elimination). The v2.1 pilot received $N=346$ usable student responses across 29 courses at 28 institutions.

When results came back, we discovered that we had inadvertently made the problem nearly impossible: only 3\% of students answered correctly! Performance of prior versions of the item, coupled with our experience as educators, led us to suspect that we were once again measuring something other than students' conceptual understanding. Discrimination was surprisingly high for an item of this difficulty, $\rho^*=0.19(5)$, indicating that correct answers represented particularly-attentive high achievers (not guessing).

Once again, a response curve (Fig.~\ref{fig:v2.1-stats}) suggested a possible culprit: students were extremely reticent to answer ``e'' for part (i), even if they answered part (ii) correctly. Think-aloud interviews confirmed that students who initially rejected all of a-d on part (i) preferred to second-guess themselves rather than choose the disfavored multiple-choice idiom ``none of these.'' For a second time, we believe that test-taking strategy overrode genuine conceptual reasoning.

\subsection{QCCS version 2.2 (spring/fall 2025)}

For the final pilot rounds, we considered dropping Item 15 entirely due to poor performance, or else keeping only subpart (ii). However, the centrality of phase kickback to so many quantum algorithms, alongside concern over assessment objective coverage, led us to try one further round with revisions. This time, we added an option ``the state CANNOT be written as a single-qubit ket.'' This version of the survey received $N=777$ usable student responses from 55 courses and 46 institutions.

\begin{figure}[!htbp]
    \centering
    \begin{subfigure}[t]{0.387\textwidth}
        \fbox{\includegraphics[scale = 0.185, clip = true, trim = 55pt 250pt 85pt 1050pt]{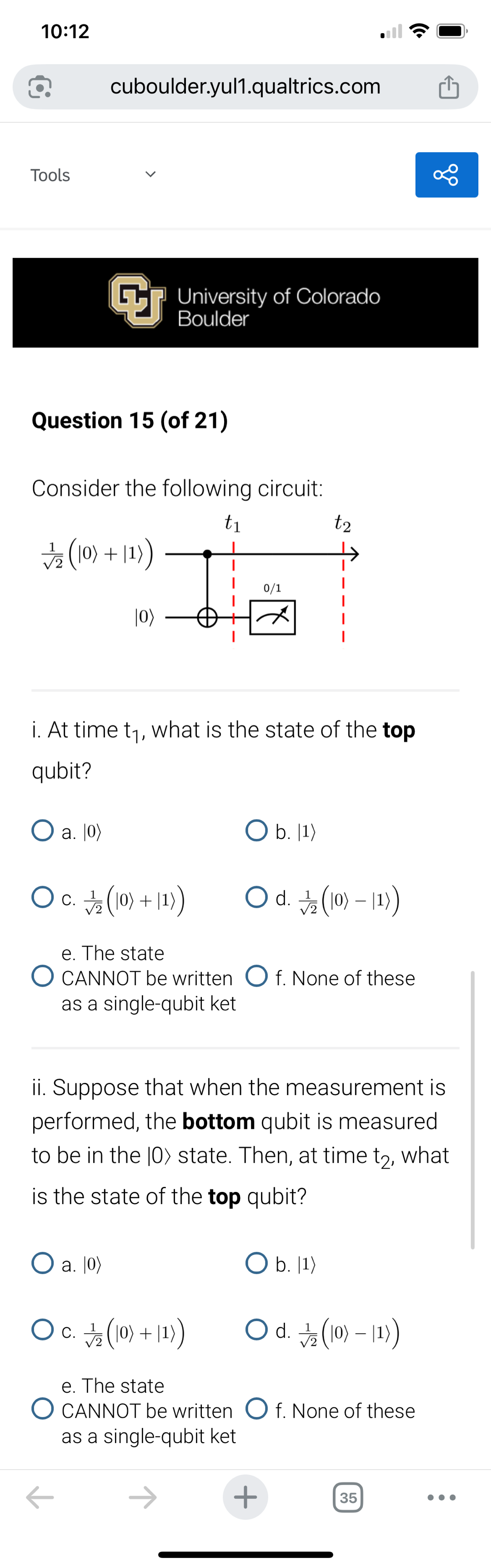}}
        \caption{Item 15 from QCCS version 2.2. Item was scored as correct if student answered ``e'' for part (i) and ``a'' for part (ii); no partial credit awarded if only one subpart answered correctly.}
        \label{fig:v2.2}
    \end{subfigure}
    \hfill
    \begin{subfigure}[t]{0.607\textwidth}
        \fbox{\includegraphics[scale = 0.385, trim = 10pt 650pt 15pt 50pt, clip = true]{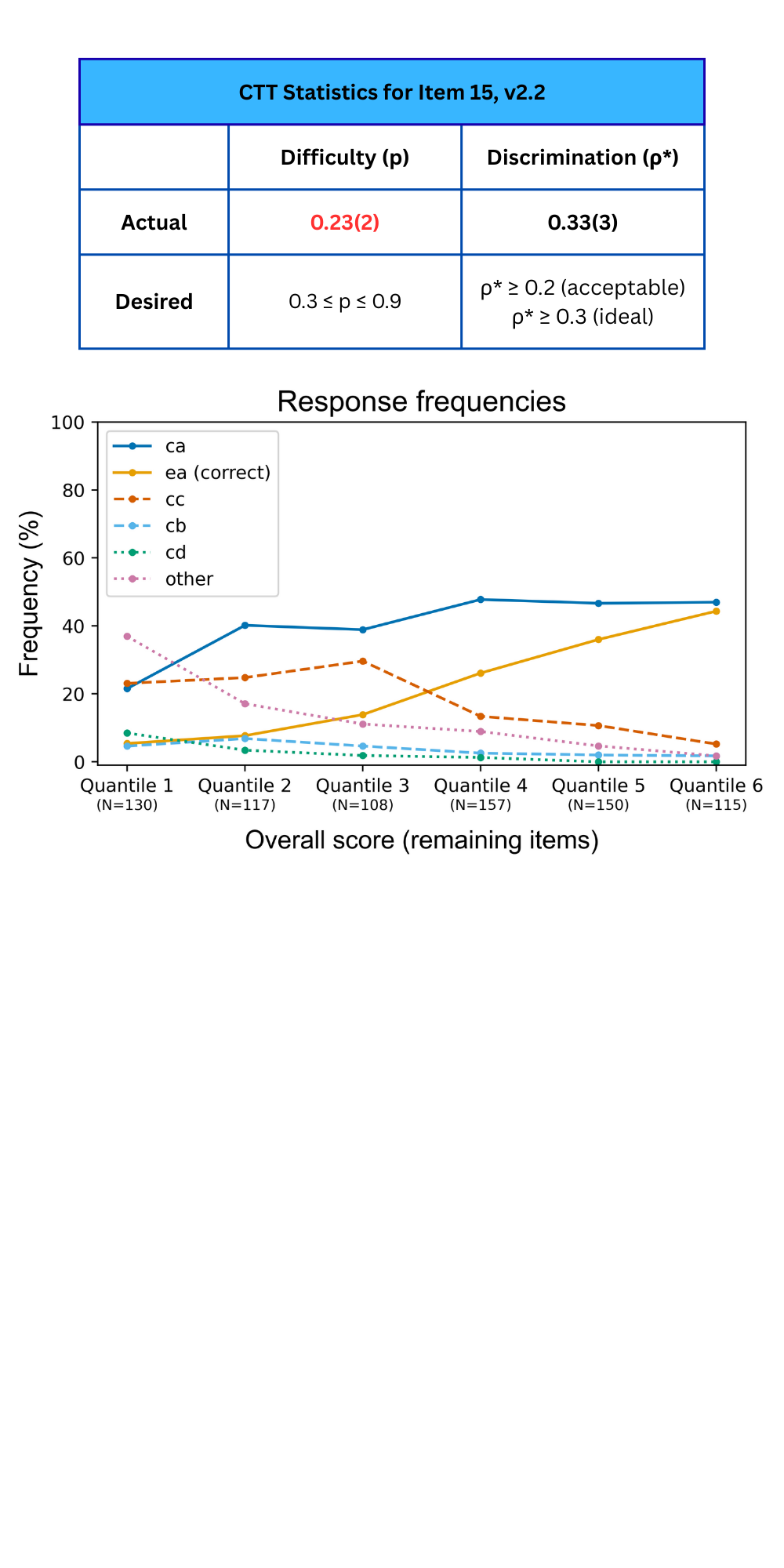}}

        \caption{Item statistics for Item 15, QCCS version 2.2. While this is still a very challenging item for students as anticipated, the good discrimination and monotonically increasing slope for the correct answer indicate a properly functioning item.}
        \label{fig:v2.2-stats}
    \end{subfigure}
    \caption{Piloted version of Item 15 during the 2025 pilot semester (v2.2).}
    \label{fig:v2.2-with-stats}
\end{figure}

When results came back from v2.2, we found that Item 15 was performing adequately. While difficulty was low ($p=0.23(2)$), performance remained far above the guess floor. The discrimination was strong ($\rho^* \gtrsim 0.3$), lending credence to our belief that the item was indeed simply measuring a hard set of skills. Specifically, to answer correctly, students must (i) conceptually understand that the CNOT gate can produce an (entangled) state not factorable as single-qubit kets and (ii) properly interpret the effect of partial measurement on an entangled superposition state.
The choice of incorrect answers aligned with our intuition about students' incorrect conceptions, lending itself to a straightforward interpretation:

\begin{itemize}
    \item The correct answer combination, ``ea,'' requires the integration of several concepts, with even relatively high achievers having a below-50\% of answering the problem correctly. However, success probability increases steadily with overall score (Fig.~\ref{fig:v2.2-stats}).

    \item The most common incorrect answer, ``ca,'' is consistent with the incorrect notion that the state of a single qubit can always be written as a ket. However, students who answer ``ca'' may recognize that measurements on an entangled state can affect both qubits. This answer peaks in popularity among above-average students, as expected for an incorrect answer still showing a degree of sophistication.

    \item The answer ``cc'' might reflect the naive conception documented in Ref.~\cite{Kushimo:2023} that the control qubit of a CNOT gate always remains unchanged; therefore, there is no mechanism for the measurement of the lower qubit to produce any back-action on the top qubit. This idea might stem from a misleading analogy to the classical XOR gate. As expected, this difficulty is most pronounced among below-median students.

    \item All other answer combinations lie near or below the guess floor.
    
\end{itemize}

Based on the totality of our analysis, we opted to retain the revised Item 15 in the final version of QCCS, finally convinced that the item measures what we intend to measure. 

\section{Discussion}
\label{sec:discussion}

Item 15 was designed to target 4 concrete assessment objectives in the context of phase kickback. Multiple early attempts failed. Yet in the process of refining the item to achieve these objectives, we uncovered several other useful findings. Among them:

\begin{itemize}

    \item \textbf{Findings about student understanding of quantum states:} Item refinement brought to light a number of insights regarding student learning. For instance, some students appear to conflate the quantum state itself with measurement probabilities \cite{Meyer:2021PERC,Plueger:2026}. Others appear to be overgeneralizing behavior of circuits from basis states alone -- suggesting further investigation is needed into student interpretation of circuit diagrams and their salient features.
    
    \item \textbf{Findings about course notational conventions:} Not all courses have the same definition of measurement. Some courses state that measurement is meaningless without specifying a basis; others implicitly treat measurement as always occurring in a specific ($Z$) basis. 
    Educators and curriculum developers should be mindful that both notational conventions coexist -- even where conventions are consistent within a course, students might reference online materials or have exposure from prior courses using the opposite convention. A word of caution to students may be warranted, as is commonly done elsewhere, incompatible notational conventions are frequently encountered (e.g.\ conflicting definitions of $\theta$ and $\phi$ in spherical coordinates in physics vs.\ math \cite{Weisstein:2025}). Where curricular materials are intended to be reusable, a simple ``0/1'' above the measurement icon appears to adequately cue students accustomed to either convention to treat the measurement as in the standard ($Z$) basis.

    \item \textbf{Importance of iterative test validation with actual students:} Even experienced test (and curriculum) developers cannot always anticipate how a question is going to perform with actual students. While disciplinary expertise and teaching experience is certainly helpful, it cannot fully substitute for ongoing field testing particularly for research-grade instruments.
    
    \item \textbf{Findings about test item structure:} Students appeared extremely reticent to answer ``none of these'' for test items even when they felt all other choices were wrong. This appears to be an ingrained test-taking habit that can confound attempts to probe student reasoning, which multiple-choice item developers should be conscious of. Similar issues are documented elsewhere in the psychometric literature \cite{DiBattista:2014}.
    
\end{itemize}

Indeed, while our manuscript addresses the quantum education community given the timeliness of assessment in our field, we anticipate the latter two findings will be of relevance to assessment developers more broadly as well as those educators who use research-based assessment materials in their classrooms. The importance of validation with actual students -- a fact broadly known within the psychometrics community, but which can come as a surprise to instructors and researchers new to research-based assessment -- cannot be overstated.

For interested readers seeking to dive deeper into the topic of assessment, we recommend starting with Refs.~\cite{Engelhardt:2009} and \cite{Madsen:2017b}.\footnote{While originally written for physics faculty, we anticipate both resources should be reasonably approachable for most of the \textit{EPJ Quantum Technology} readership. The examples drawn are primarily from introductory mechanics and electromagnetism, courses taken by most engineers and computer scientists at some points in their undergraduate studies.} These accessible resources provide detailed overviews of best practices for assessment development and administration, forming a beginner-friendly bridge to the broader psychometrics and assessment literature.

\section{Conclusions}

We have presented the saga of QCCS Item 15 not only to tell an interesting story, but to elucidate the process of developing a good QIS assessment item and what can be learned along the way. In education research, behind every research-based tutorial, assessment instrument, or claim about student thinking is usually a years-long journey refining interview protocols or assessment instruments (seldom reported publicly). Often, as in all science, important discoveries are made by accident. 

As Item 15 illustrates, student response patterns to a seemingly straightforward physics problem can be baffling to interpret. Changes that appear minor by eye, such as between versions 2.1 and 2.2, can dramatically affect performance. Test items or interview questions may appear foolproof in whiteboard discussions among researchers, then fail spectacularly among actual students for mundane reasons like ingrained test-taking strategies. Yet iterative item refinement is itself highly informative, often revealing answers to important unasked questions. 

Above all, we are reminded by this work that experts, even DBER researchers, are seldom able to anticipate in advance how students actually think about quantum computing. Curriculum and assessment development for the quantum workforce is inescapably a process of trial-and-error; fortunately, there is much we can learn from our successes and failures alike along the way.
%TC:ignore

\section{Abbreviations list}
\begin{itemize}
\item \textbf{DBER:} Discipline-based education research
\item \textbf{QCCS:} Quantum Computing Conceptual Survey (research-based assessment instrument, see Ref.~\cite{Meyer:2025QCCS}).
\item \textbf{QIS:} Quantum information science
\end{itemize}

\section{Declarations}

\textbf{Data availability:} The datasets analyzed during the current study are not publicly available due to human subjects privacy considerations but will be made available in deidentified form from the corresponding author upon reasonable request.

\textbf{Competing interests:} The authors declare that they have no competing interests.

\textbf{Funding:} This work was supported by the University of Colorado Boulder Department of Physics, the California State University-Fullerton Department of Physics, the NSF Graduate Research Fellowship Program, and NSF Grants Nos.\ 2011958, 2012147, and 2143976.

\textbf{Human subjects data:} All data collection, including both interviews and large-scale pilot testing, was by informed consent and approved through the University of Colorado Boulder institutional review board under protocol 20-0583.

\textbf{Author contributions:} GP conceived of the concept for this manuscript, with input from all authors. JCM performed the majority of data analysis and writing of the manuscript. All authors were closely involved with the development and validation of the QCCS (including, but not limited to, Item 15) and read and approved the final manuscript.

\textbf{Acknowledgments:} We thank Soren Rohi-Plueger and Michael Burnes for assistance with interviews and analysis thereof, including proposing the modified form of Item 15 for version 2.1 that closely approximated the final product. We also thank Molly Griston for contributing to statistical analysis.

\bibliography{ref.bib}

\end{document}